\documentstyle[aps,prc,epsf]{revtex}
\begin{document}
\title{Short-range correlations in nuclear matter using Green's functions
within a discrete pole approximation}
\author{Y. Dewulf, D. Van Neck and M. Waroquier} 
\address{Laboratory for Theoretical Physics, University of Ghent,
Proeftuinstraat 86, B-9000 Gent, Belgium}
\maketitle
\vspace*{.5cm}
\begin{abstract}
We treat short-range correlations in nuclear matter, induced by the repulsive core
of the nucleon-nucleon potential, within the framework of self-consistent Green's 
function theory. The effective in-medium interaction sums the ladder diagrams
of both the particle-particle and hole-hole type. The demand of self-consistency 
results in a set of nonlinear equations which must be solved by iteration. We explore the 
possibility of approximating the single-particle Green's function by a limited number of
poles and residues.
\end{abstract}
\vspace*{1cm}
\newline 
keywords: nuclear matter, Green's function
\vspace*{.5cm}
\section{Introduction}
The correct treatment of short-range correlations when 
performing nuclear calculations using realistic NN-interaction 
is of great importance \cite{mutherpolls}. Realistic NN-interactions are fitted to
the two-nucleon scattering data and the deuteron binding energy 
and contain a large repulsive core for small internuclear distances \cite{stoks}. 
When this interaction is used in quantum-mechanical perturbation theory,
convergence is no longer guaranteed as the dominant contribution of
the repulsive core will lead to large and repulsive two-body matrix elements. 
A Hartree-Fock calculation starting from a realistic interaction will
lead to an unbound system, and the energy of the ground state will be
highly dependent on the specific potential used \cite{polls}.

Within the self-consistent Green's function (SCGF) formalism short-range correlations
are treated by replacing the realistic interaction with a medium-modified
effective interaction. This effective interaction 
is obtained by summing all diagrams
of the ladder type. 

Short-range correlations will lead to a depletion
of the hole states and a partial occupation of the particle states in
the ground state of the system. This deformation of the Fermi sea 
will reduce the pairing correlations \cite{bozek} and can lead to a 
non-superfluid ground state at the empirical saturation density.
Furthermore an additional density dependence is introduced,
which influences the saturation properties
of nuclear matter. 

A self-consistent treatment employing the full energy dependence of 
the spectral function is feasible for nuclear matter at finite temperatures
\cite{bozek2}, but becomes very difficult at zero temperature, due to the
complicated structure of the spectral function.
Therefore a number of approximations for the spectral function 
have been developed in the recent past \cite{ramos,vonderfecht,dejong,gearhart1,gearhart2,roth}.
In this work we propose a discrete pole approximation for the
Green's function. Using this approximation the deformation of the
Fermi sea is incorporated in the evaluation of the effective interaction.
Different discretization schemes are possible. Here the BAGEL scheme has been 
used to generate the exact location and strength of the discrete poles.
Results are displayed for a modified Reid interaction.

\section{Formalism}
Green's functions constitute a useful tool for investigating the
properties of correlated many-particle systems. 

When considering infinite nuclear matter, the symmetries of the system
will seriously reduce the degrees of freedom and the 
single-particle Green's function will be completely 
determined by the magnitude of the momentum $p$ and the energy of the 
particle $\omega$.
This propagator can then be represented in Lehmann's representation as,
\begin{equation}
g(p,\omega)=
\int_{-\infty}^{\epsilon_{F}} d\omega\,'
\frac{S_{h}(p,\omega')}{\omega-\omega\,'-i\eta} 
+\int_{\epsilon_{F}}^{\infty} d\omega\,'
\frac{S_{p}(p,\omega')}{\omega-\omega\,'+i\eta}.  
\label{eq:lehmann}
\end{equation} 
The particle spectral function $S_{p}(p,\omega)$ represents the probability
of adding a particle with momentum $p$ to the system, while leaving the
resulting system with an excitation energy $\omega$. The hole spectral
function $S_{h}(p,\omega)$
is related in a similar way to the removal of a particle with
momentum $p$.

\begin{figure}[fig1]
\centering{
\epsfbox{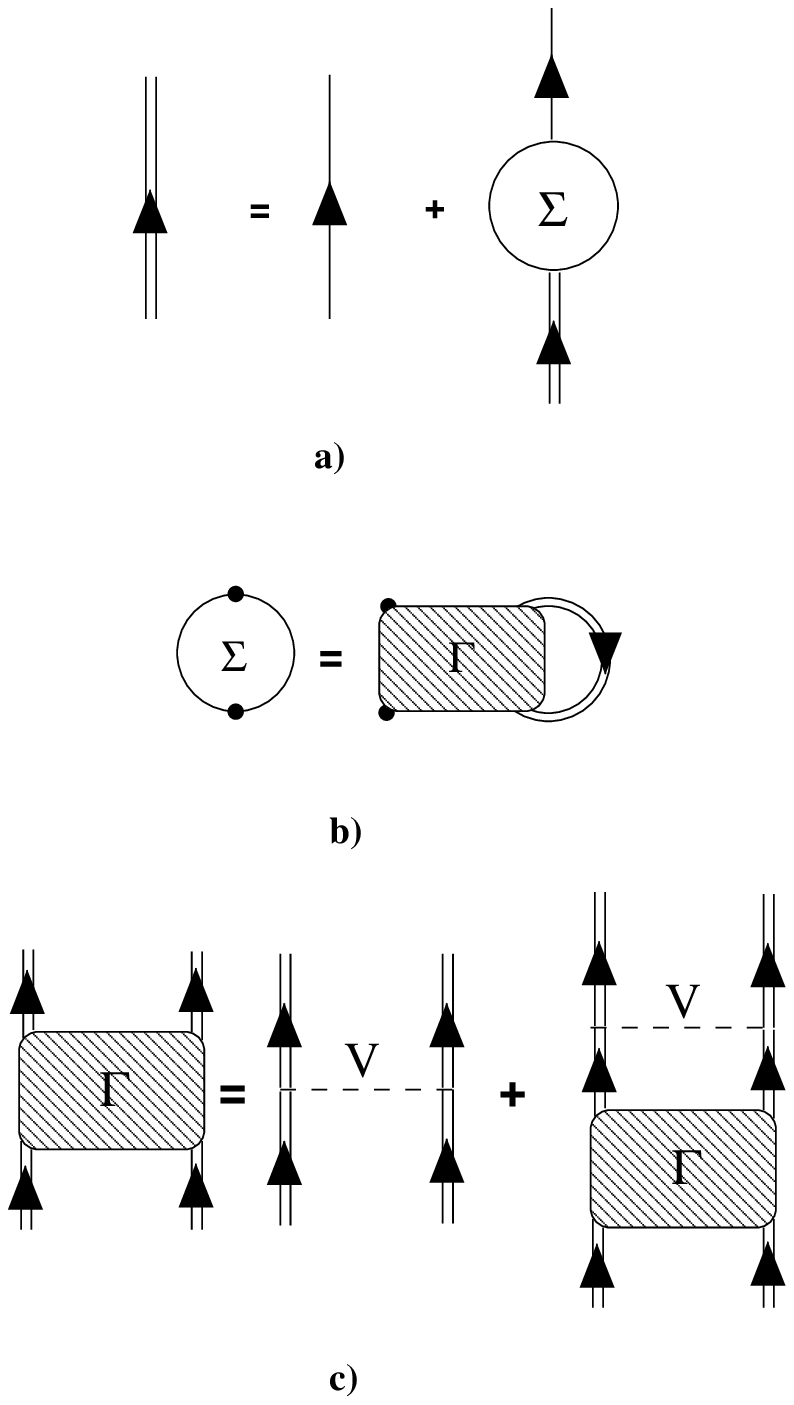}
\caption{Diagrammatic representation of a) the Dyson equation
b) the self-energy and c) the effective interaction in the ladder
approximation. }
\label{fig:selfconsist}}
\end{figure}

Using the Dyson equation, the correlated spectral functions can be determined, 
\begin{equation}
\begin{array}{ll}
S_{h}(p,\omega)=&
\left\{
\begin{array}{ll}
\frac{1}{\pi}\frac{\mbox{Im}\,\Sigma(p,\omega)}
{(\omega-p^2/2m-\mbox{Re}\,\Sigma(p,\omega))^{2}+
(\mbox{Im}\,\Sigma(p,\omega))^{2}}& 
\mbox{if}\; \omega < \epsilon_{F} \\
0& 
\mbox{if}\; \omega > \epsilon_{F},
\end{array}
\right. \\
\\
S_{p}(p,\omega)=&
\left\{
\begin{array}{ll}
0& \mbox{if}\; \omega < \epsilon_{F} \\  
-\frac{1}{\pi}\frac{\mbox{Im}\,\Sigma(p,\omega)}
{(\omega-p^2/2m-\mbox{Re}\,\Sigma(p,\omega))^{2}
+(\mbox{Im}\,\Sigma(p,\omega))^{2}}
& \mbox{if}\; \omega > \epsilon_{F} \\
\end{array}
\right. .
\end{array}
\label{eq:specfull}
\end{equation}
The irreducible self-energy $\Sigma$ is a complex quantity and
represents the interaction with the other particles. 
Because of the short-range correlations induced by the repulsive
core of a realistic NN-interaction, the self-energy must be expanded
in terms of an energy-dependent effective interaction $\Gamma$,
\begin{equation}
\Sigma(p,\omega)=
\int\frac{d\vec{p}\;'}{(2\pi)^3} 
\int \frac{d\omega'}{2\pi i}
\langle \vec{q}\;|\Gamma(P,\omega+\omega')|\vec{q}\;\rangle
g(p',\omega'),
\label{eq:selfen}
\end{equation}
with $\vec{P}=\vec{p}+\vec{p}\;'$ and $\vec{q}=(\vec{p}+\vec{p}\;')/2$.
The summation of both the spin and isospin variables is suppressed for
notational convenience. A diagrammatic representation of the self-energy 
is given in Fig.\ref{fig:selfconsist}.

The effective interaction $\Gamma$ is
calculated from the realistic NN-interaction by summing the infinite
subset of ladder diagrams. In contrast to the Brueckner-Hartree-Fock
(BHF) approach, SCGF retains 
both particle-particle (pp) and hole-hole(hh)
propagation in the intermediate two-particle propagator \cite{ramos},
\begin{eqnarray}
&&\langle \vec{p}|\Gamma(P,\Omega)|\vec{p}\;'\rangle
=\langle \vec{p}|V|\vec{p}\;'\rangle  
+\int \frac{d\vec{q}}{(2\pi)^{3}} \langle \vec{p}|V|\vec{q} \rangle
\nonumber \\
&& \hspace*{1cm}
\times\left[
\int d\omega \int d\omega'\frac{
S_{p}(p_{1},\omega)S_{p}(p_{2},\omega')}
{\Omega-\omega-\omega'+i\eta}
-\int d\omega\int d\omega'\frac{S_{h}(p_{1},\omega)
S_{h}(p_{2},\omega')}{\Omega-\omega-\omega'-i\eta}
\right]
\langle \vec{q}|\Gamma(P,\omega)|\vec{p}\;'\rangle,
\label{eq:effective}
\end{eqnarray}
with $\vec{p}_{1,2}=\vec{P}/2\pm\vec{q}$.

Self-consistency becomes a crucial requirement if one wants
to fulfill certain elementary conservation laws such as the conservation
of the number of particles.
Therefore the spectral function obtained from Eq.(\ref{eq:specfull}) should be
used in the evaluation of the self-energy, Eq.(\ref{eq:selfen})
and the effective interaction, Eq.(\ref{eq:effective}).
The resulting set of coupled, non-linear equations is solved by
iteration. Starting from an initial guess for the Green's function,
one calculates the self-energy and the effective interaction. The 
Dyson equation then leads to a Green's function different from 
the original one. This Green's function can again be used in the
evaluation of the effective interaction and the self-energy. This procedure
continues until both self-energy and Green's function have converged.

At $T=0$ (zero temperature), the evaluation of the effective 
interaction through the successive iterations is a cumbersome task. The 
combination of very sharp peaks as well as a 
broad background distribution in the spectral function makes an
accurate numerical evaluation difficult.
It is therefore of importance to look for suitable 
approximations for this spectral function.
These approximations should retain the most prominent features of the full
spectral function, Eq.(\ref{eq:specfull}), but at the same time they must allow
a fast evaluation of the self-energy  
and the effective interaction.
\section{Approximated spectral functions}
The most basic approximation for the spectral function 
is obtained by neglecting the imaginary
part of the self-energy, so that the full spectral function of 
Eq.(\ref{eq:specfull}) is replaced by a single delta peak at the 'on-shell energy',
\begin{equation}
\epsilon_{qp}(p)=p^{2}/2m+
\mbox{Re}\Sigma(p,\epsilon_{qp}(p)).
\end{equation} 
This approach corresponds to an extension of the Brueckner-Hartree-Fock(BHF)
prescription, 
including hh-propagation in the effective interaction, and
is usually referred to as the quasi-particle scheme. Extended calculations
within this scheme were carried out for both model interactions \cite{ramos}
and realistic NN-interactions \cite{vonderfecht}. In the latter calculation \cite{vonderfecht}
the on-shell energy is replaced by the mean removal energy for hole states.
As a result a finite gap in the energy spectrum is obtained at the 
Fermi energy. This gap is necessary to avoid the pairing instability,
that occurs when hh-propagation is included within a quasi-particle scheme.

A major drawback of the quasi-particle approach is the neglect
of the broad background distributions in both particle and hole spectral
function. These background distributions are assumed to carry 
about 15-25\% of the total strength \cite{mutherpolls} and are responsible
for a serious deformation of the Fermi-sea. The incorporation of
the effects of this deformed Fermi-sea in the calculation of the
effective interaction leads to an additional density dependence, that contributes
to a full description of the saturation mechanism of
nuclear matter.

In recent years several attempts have been undertaken to
perform self-consistent calculations using a spectral function more
closely related to the full spectral function. The group of Dickhoff
parametrizes the spectral function in each
iteration \cite{gearhart1,gearhart2,roth}. 
In this way the integrals can be carried out in an analytic way.
De Jong \cite{dejong} uses a mixed approach, in which for $p<k_{F}$
the hole spectral function is replaced by a renormalized delta peak, whereas
the complete energy dependence of the 
particle spectral function is retained. The integration of this
particle spectral function does not lead to large difficulties, since
there is no sharp quasi-particle peak in the particle spectral function
for $p<k_{F}$.
For states with $p>k_{F}$
a similar mixed approach is then used.

In this paper we propose an extension of the quasi-particle scheme,
that aims at incorporating the off-shell propagation of the particles.
The Green's function is represented by a small set of discrete 
poles, 
\begin{equation}
g(p,\omega)=\sum_{i}\frac{f_{i}(p)}{\omega-F_{i}(p)+i\eta}
+\sum_{j}\frac{b_{j}(p)}{\omega-B_{j}(p)-i\eta}.
\label{eq:disgreen}
\end{equation}
The poles labeled $B_{i}$ are located below the Fermi energy, while
those labeled $F_{i}$ are situated above this energy. 
The poles and their
corresponding residues are 
chosen in such a way that they form a reasonable approximation
of the fully dressed Green's function. In addition to a peak corresponding
to the quasi-particle peak, additional peaks are used to simulate the
background distribution for both positive and negative energies. 

Several discretization schemes for the spectral function can be proposed. 
For each scheme one should obtain poles $F,B$ and residues
$f,b$ that are continuous functions of the momentum $p$. Furthermore one
wants to recover the complete spectral function in the limit of 
an infinite number of poles. 
The specific discretization scheme used in this paper is based on the
BAGEL approach, which has proven to
be very successful in the description of the long-range correlations in
finite nuclei \cite{amir1,muther5,muther1,dewulf}.

The poles and residues obtained within the BAGEL scheme
reproduce the correct lowest order energy-weighted moments of 
the continuous spectral function at each iteration,
\begin{eqnarray}
m_{k}(p)=\sum_{i}{f_{i}(p)}(F_{i}(p))^{k}
+\sum_{j}{b_{j}(p)}(B_{j}(p))^{k} 
=\int_{-\infty}^{+\infty}\omega^{k}S(p,\omega)d\omega,\;\;\;\forall p.
\label{eq:moments}
\end{eqnarray}
The BAGEL algorithm allows for a direct calculation of the poles and the
residues of the Green's function from the imaginary part of the self-energy,
without having to calculate the integral 
on the right-hand side of Eq.(\ref{eq:moments}).  
For an approximation using $D$ poles, one expects the moments up to
$2D-1$ to be reproduced. However, due to the large asymmetry in the
self-energy, the BAGEL algorithm must be applied separately
to both the forward ($\omega>\epsilon_F$) and backward 
($\omega<\epsilon_F$) part of the self-energy.
Therefore only moments $k \leq D$ will be reproduced.

\section{Results}
We first display results for the most simple extension of
the quasi-particle scheme: a three pole BAGEL approximation.
A modified version of the Reid93-interaction \cite{stoks} was used for these results. 
In order to avoid pairing instabilities, the $^1S_0$ and the
$^3S_1-^3D_1$ partial waves of this interaction are multiplied by a factor 0.75 and 0.5 
respectively. 

This pairing instability leads to the appearance of complex poles 
\cite{dickhoff} in the
effective interaction, which makes a calculation along the lines of the
scheme sketched out
above impossible. However a self-consistent calculation, using the
full energy dependence of the spectral function might lead to the disappearance
of these pairing instabilities. In this case the projection operators
used in the calculation of the effective interaction relate to
a Fermi sea that
is deformed due to short-range correlations, rather than to the uncorrelated
Fermi sea. This
causes a reduction of the effective interaction for energies close to 
$2\epsilon_{F}$ and will also reduce the pairing correlations
\cite{bozek}. Such a mechanism
can lead to a normal ground state at the empirical saturation density
($k_{F}=1.36 fm^{-1}$) instead of a superfluid ground state, provided the
short-range correlations are strong enough.
As a consequence a self-consistent procedure using the full realistic interaction
would become feasible, without having to treat the possible pairing of nucleons. 
The discrete pole approximation is a suitable tool to incorporate the
effects of the deformed Fermi sea in the calculation of the effective
interaction. If the specific discretization scheme fails to include the complete
effects of the correlated Fermi sea, the pairing instability
remains unless a reduced interaction is used. When using such a modified
interaction, one cannot expect to reproduce the correct empirical values
of the binding energy and the occupation probabilities, but the results nevertheless hint at 
the new mechanisms at work when off-shell propagation is included in the calculation             .

\begin{figure}[fig4]
\centering{
\epsfbox{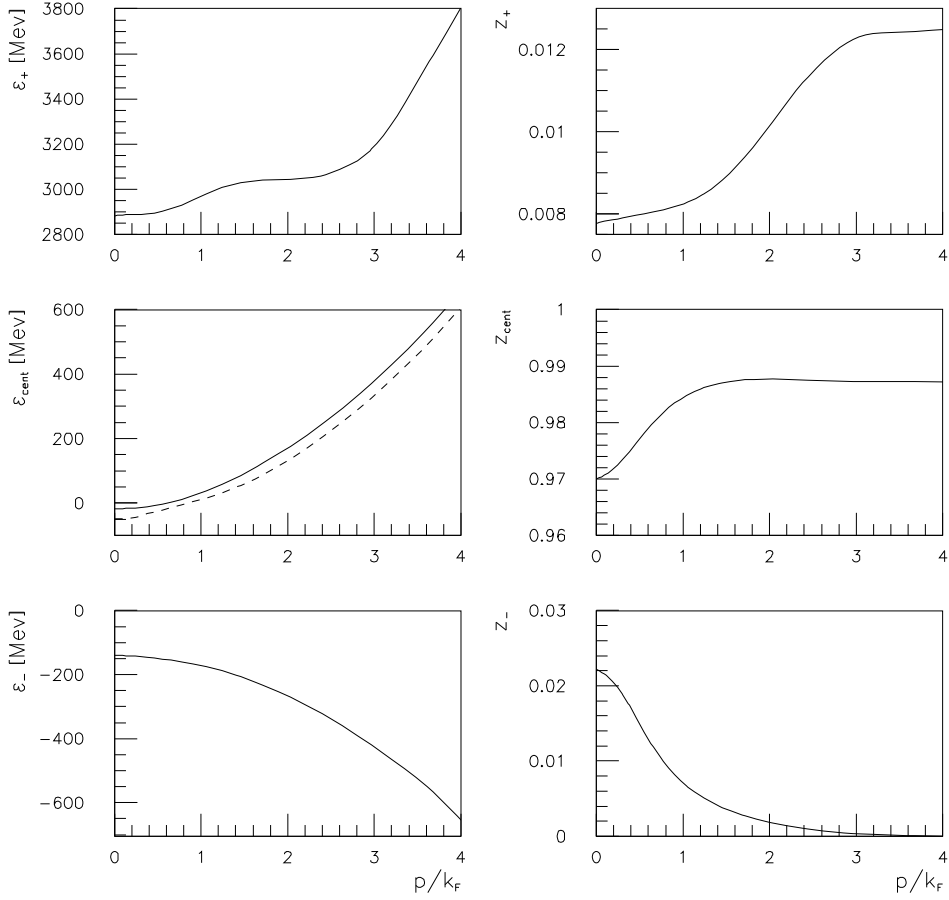}
\caption{A three-pole BAGEL spectrum after the convergence of the scheme
has been reached ($k_{F}$=1.36 fm$^{-1}$).The left-hand
side displays the location of the poles, while the right-hand side
shows the corresponding residues. In the central plot on the left
the full line corresponds to the central BAGEL pole, 
while the dashed line corresponds to the on-shell
energy corresponding with the self-energy obtained  within the same BAGEL approximation.}
\label{fig:bagelspec}
}
\end{figure}

\begin{figure}[fig5]
\centering{
\epsfbox{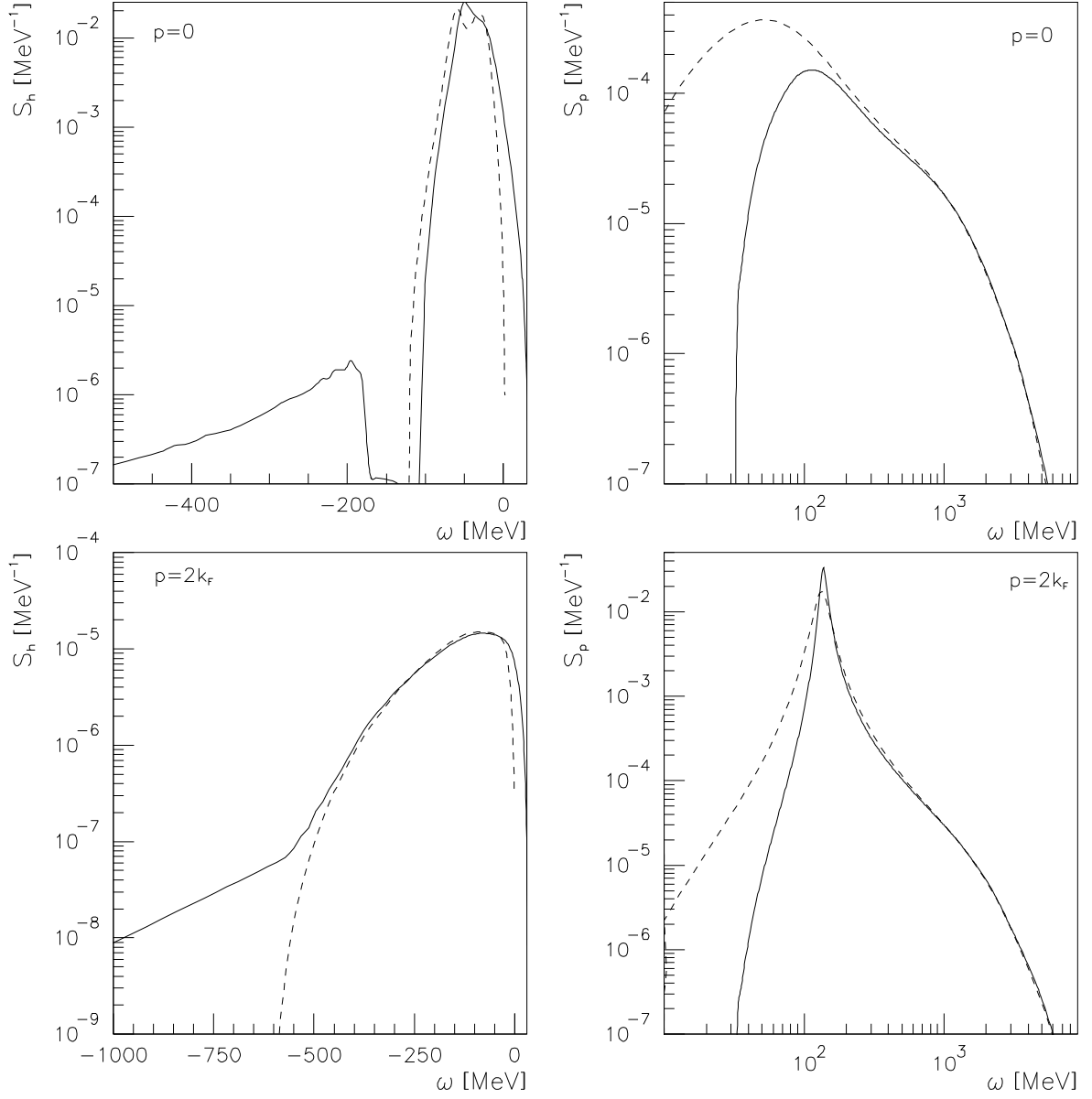}
\caption{The hole and particle spectral function calculated using
Eq.(\ref{eq:specfull}) after the convergence of the BAGEL spectrum,
at $k_F$=1.36 fm$^{-1}$.
The upper plots correspond to a momentum $p=0$, the lower plots 
to $p=2k_{F}$. Note the use of a logarithmic energy scale for
the particle spectral functions, illustrating the large spreading
of particle strength due to the short-range correlations. 
(Also the linear energy scale used for the right-hand side plots
differs for both momenta). The full line corresponds to a BAGEL
calculation, the dashed line to the dressed quasi-particle result.}
\label{fig:specfunc}
}
\end{figure}

Fig.\ref{fig:bagelspec} displays the poles and the residues of a 
three-pole BAGEL scheme as a function of the momentum. 
\begin{itemize}
\item The central 
pole (labeled ``c'') has a behaviour similar to the quasi-particle pole, but
carries more than 97$\%$ of the strength for all momenta. 
This is considerably more than 
the strength carried by the quasi-particle pole, which is  
typically 65-75$\%$ for hole states \cite{mutherpolls}. Furthermore this central pole 
is located about 20 MeV above the on-shell energy for all momenta
(dotted line). 
These observations 
indicate that in the BAGEL scheme the central pole will not only
represent the quasi-particle pole, but will also contain a considerable
fraction of the forward background distribution. 
\item The remainder
of this background distribution is represented by a pole located
at high positive energies (labeled ``+''). 
This behaviour is typical for nuclear matter calculations using a realistic force.
The large value for the energy
of the ``+''-pole is a result of the
the short-range correlations induced by the repulsive
core of the interaction. The exact location of this pole
depends strongly on the specific interaction used.

\item Finally the background distribution 
at negative energies is represented by the pole labeled ``-'', whose
location equals roughly 
\begin{equation}
B_{-}(p)\sim-p^{2}/2m \nonumber.
\end{equation}
This pole 
corresponds to the high-momentum components in the nuclear
wave function \cite{ciofi}.
\end{itemize} 
The full energy dependence of the spectral function 
calculated from Eq.(\ref{eq:specfull}) is 
displayed in Fig.\ref{fig:specfunc} after convergence, in a quasi-particle
scheme (dashed line) and in a three-pole BAGEL scheme (full line), for two
momenta.
The most pronounced difference is the energy range where the spectral
function is non-vanishing. In the quasi-particle calculation the hole
spectral function will only be different from zero above a threshold energy.
This is no longer the case in a BAGEL calculation, where the hole spectral
function extends to $-\infty$. One should also note the shell-like
structure of the BAGEL spectral function. This behaviour is a result of
the discrete pole approximation for the propagators: the
different energy intervals leading to a non-vanishing spectral function 
can be attributed to different combinations of the discrete 
poles contributing to the effective interaction and the self-energy. 

The difference in spectral function close to the Fermi energy should
be ascribed to the different locations of the central BAGEL pole 
and the quasi-particle energy. 
This deviation also causes a number of
ambiguities when evaluating the occupation probability
and binding energy. 

The occupation probability is shown in Table \ref{tab:occupation}.
Results are displayed for three different calculations. 
The column labeled ``BAGEL'' corresponds to the sum of the backward
BAGEL poles. This is the occupation probability used in the evaluation
of the effective interaction within the BAGEL scheme. Due to the large amount 
of strength concentrated in the central pole of the BAGEL spectrum, 
the induced depletion is not large enough to avoid the appearance of a 
pairing instability when the full interaction is used.
The other columns in Table \ref{tab:occupation}
correspond to the integration of the full hole-spectral function obtained after
convergence of the quasi-particle scheme and the three-pole
BAGEL scheme. 
When comparing both dressed approximations we observe that the quasi-particle
scheme leads to a larger depletion than the BAGEL scheme. The underlying reason
is the feedback of the depleted Fermi sea on the effective interaction. The
BAGEL scheme incorporates the effects of this correlated Fermi sea to some 
extent, which causes a reduction of the effective interaction around
$2\epsilon_{F}$. These preliminary calculations suggest that the inclusion of
off-shell propagation leads to a reduction of the depletion.
It should be noted that the BAGEL approximation does not conserve the total number
of particles, since one finds a particle excess of $4\%$ using the occupation 
probabilities obtained by summing the discrete poles and an excess of $8\%$ using
the integrated hole spectral function. This is again caused by the specific
features of the BAGEL scheme and such a behaviour is not expected to show up
when other discretization schemes are used. One should note that these
occupation probabilities were obtained using a reduced interaction. 
Calculations with a full interaction should yield a larger depletion. 
Both a quasi-particle calculation using the soft-core Reid interaction 
\cite{vonderfecht} and
a CBF-calculation using the Urbana V$_{14}$ potential \cite{benhar}
predict an occupation probability of $n(0)=.82$.

The problems associated with the position and strength of the central pole,
cannot be avoided within a BAGEL scheme. When larger
sets of BAGEL poles are included in the description of the Green's function,
the central pole after the first iteration will show a closer resemblance 
with the quasi-particle pole as is shown for $p=0$
in table \ref{table:tablemb1}.
Table \ref{table:tablemb2} displays the second and third iteration of
a five-pole calculation. One can see that already 
in the second iteration, the feedback-mechanism 
embedded in the self-consistent
calculation will reattribute part of the strength of the outward
poles to the central pole, which at the same time 
shifts to higher energies. Hence, the self-consistent inclusion 
of a larger set of BAGEL poles does not improve the correspondence 
between the central BAGEL pole and the quasi-particle behaviour.

One can try to alleviate the problems 
caused by the location of the central BAGEL pole by 
performing a shift of the central BAGEL pole or of
the complete BAGEL spectrum. However, such a scheme does not converge,
and due to this shift
the correct lowest order moments of the spectral function are no longer 
reproduced. 
Furthermore this shift does not alter the strength attributed 
to the central pole so that the pairing-instability is not avoided
when the full realistic interaction is used.

It should be clear at this point that if one wants to
benefit from the advantages of the discrete pole scheme, 
a new discretization scheme should be devised.
When constructing this scheme, one should
focus on a better description of the quasi-particle behaviour, 
by fixing the location of the central-pole at the quasi-particle
energy. The other poles can be
used to describe the features of the extended background distribution,
including the sumrules of the spectroscopic strength.
Work along these lines is currently in progress.

\section{Summary}
The discrete pole approximation proposed in this letter 
is an extension of the quasi-particle scheme, incorporating the
 off-shell propagation of the nucleons.
This off-shell propagation is absent in a one-pole scheme, since 
the background distributions are neglected within such a scheme.
Using a multiple-pole approximation for the Green's function, 
the deformation of the Fermi sea, originating from the short-range 
correlations is taken into account in the evaluation
of the effective interaction. 
This opens the possibility of studying the effects of self-consistency on 
pairing properties of nuclear matter. 
The deformation of the Fermi sea also introduces an 
additional density dependence in the binding-energy 
which is important for the correct description of the saturation mechanism.
Preliminary calculations using a BAGEL-discretization scheme and a model potential,
suggest a reduction of the depletion caused by a new feedback mechanism originating
from the use of a correlated Fermi sea. The BAGEL scheme however is plagued by a 
weak correspondence between the central pole and the quasi-particle pole.
The use of an improved discretization scheme should solve this problem.

\begin{table}[table:occupation]
\caption{The occupation probability calculated in three different
approaches, with $k_{F}=1.36fm^{-1}$. 
The full quasi-particle and full BAGEL results are 
obtained by integrating the respective spectral functions up to 
the Fermi energy. The BAGEL result refers to the sum of the BAGEL
poles located below the Fermi energy.}
\begin{center}
\begin{tabular} {lccc}
& full q.p.
& BAGEL
& full BAGEL
\\
\hline
$p$=0.0&0.91&0.99&0.94 \\
$p$=0.7$k_{F}$&0.87&0.99&0.92\\
\hline
$p$=1.2$k_{F}$&2.8 10$^{-2}$&5.1 10$^{-3}$& 3.4 10$^{-2}$ \\
$p$=2.0$k_{F}$&3.2 10$^{-3}$&1.8 10$^{-3}$& 3.4 10$^{-3}$ \\
\end{tabular}
\end{center}
\label{tab:occupation}
\end{table}

\begin{table}[table:bhf]
\caption{The location and strength of the central pole for $p=0$ and 
$k_{F}$=1.36 fm$^{-1}$, in the BAGEL-scheme 
after the first iteration, when larger sets of $d=3,5,\dots,13$
BAGEL-poles are included in the calculation. 
The quasi-particle energy for this momentum 
equals $-49.47 MeV$.
}
\begin{center}
\begin{tabular} {|c||c|c|}
$d$ &$\epsilon_{c}$[MeV]& $z_{c}$\\
\hline
3&-16.88&0.87\\
5&-17.24&0.83\\
7&-17.93&0.80\\
9&-18.89&0.77\\
11&-19.95&0.73\\
13&-20.97&0.67\\
\end{tabular}
\end{center}
\label{table:tablemb1}
\end{table}

\begin{table}[table:bhf]
\caption{Results of a five pole BAGEL calculation for the first 
three iterations with $p=0$,$k_{F}$=1.36 fm$^{-1}$. }
\begin{center}
\begin{tabular} {|c||c|c|}
iter. &$\epsilon_{c}$[MeV]& $z_{c}$\\
\hline
1&-17.24&0.83\\
2&-12.66&0.92\\
3&$\;\;$-9.58&0.96\\
\end{tabular}
\end{center}
\label{table:tablemb2}
\end{table}

\end{document}